\documentclass[aps,prd,amsmath,amsfonts,a4paper,11pt,reprint,twocolumn,square,numbers,showpacs,superscriptaddress,floatfix,sort&compress,nofootinbib]{revtex4-1}
\usepackage{hyperref}
\usepackage{amssymb}
\usepackage{bm}
\usepackage{natbib}
\usepackage{graphicx}
\expandafter\let\csname equation*\endcsname\relax
\expandafter\let\csname endequation*\endcsname\relax
\usepackage{amsmath}
\usepackage{color}

\RequirePackage{ifpdf}
\ifpdf
\else 
\fi

\newcommand{\para}[1]{\par\vspace{2mm}\noindent\textbf{\emph{{#1}}.---}}

\usepackage{titlesec}
\titlespacing*{\section}{0pt}{0.6cm}{0.5cm}

\newcommand{\mpl}{M_{\rm P}\,}
\newcommand{\cT}{{\cal T}}

\newcommand{\beq}{\begin{equation}}
\newcommand{\eeq}{\end{equation}}

\newcommand{\bea}{\begin{eqnarray}}
\newcommand{\eea}{\end{eqnarray}}

\begin{document}

\begin{center}
\rightline{\small DESY-18-125}
\vskip -3cm
\end{center}

	\title{Primordial Gravitational Waves and the Swampland}

	\author{Mafalda Dias}
	\email{mafalda.dias@desy.de}
	\author{Jonathan Frazer}
	\email{jonathangfrazer@gmail.com}

\author{Ander Retolaza}
\email{ander.retolaza@desy.de}
\author{Alexander Westphal}
\email{alexander.westphal@desy.de}
\affiliation{Deutsches Elektronen-Synchrotron, DESY, Notkestra\ss e 85, 22607 Hamburg, Germany}

	\begin{abstract}
The swampland conjectures seek to distinguish effective field theories which can be consistently embedded in a theory of quantum gravity from those which can not (and are hence referred to as being in the swampland). We consider two such conjectures, known as the Swampland Distance and de Sitter Conjectures, showing that taken together they place bounds on the amplitude of primordial gravitational waves generated during single field slow-roll inflation. The bounds depend on two parameters which for reasonable estimates restrict the tensor-to-scalar ratio to be within reach of future surveys. 
	\end{abstract}
	
	\maketitle

Despite the large landscape of effective field theories (EFTs) which can be derived from string theory, this set is measure zero in the set of all EFTs~\cite{Denef:2008wq}. So while the landscape of string theory may be large, it is thought to be surrounded by a truly vast swampland of EFTs which are not compatible with quantum gravity. A long standing goal has therefore been to establish necessary criteria for an EFT to not be in the swampland. 
Numerous attempts to establish these criteria have been made in the past decades from different perspectives such as black hole physics or evidence from string theory compactifications. So far these attempts have lead to a series of conjectures, known as the   weak gravity \cite{ArkaniHamed:2006dz} and swampland conjectures. 

In this letter we focus on two of these conjectures, the Swampland Distance Conjecture (SDC) and the Swampland de Sitter Conjecture (SdSC), with the objective of studying their combined implications for early universe cosmology and in particular for the amplitude of primordial tensor perturbations. 
By building on arguments from string theory compactifications and EFT, we relate the criteria from these conjectures to limits on the slow-roll parameter $\epsilon$ and, consequently, on the tensor-to-scalar ratio $r$ in single field inflation. We find that these swampland conjectures lead to a rough expectation for $r$ which can be fully accessible and tested by cosmic microwave background experiments over the next 10 years.

\para{Swampland Distance Conjecture (SDC) \cite{Ooguri:2006in,Klaewer:2016kiy}} This conjecture states that whenever  one performs a canonical displacement in moduli space $\Delta\phi  \gg   \mpl \mathcal{D} $ (where $\mathcal{D}$ is some order one constant that  depends on the details of the compactification), an infinite tower of massive states becomes exponentially light, invalidating the corresponding EFT.  
Specifically, the masses of this tower of states $\cT$ scale with the moduli displacement as 
\beq
m^2_\cT\sim \exp \left( -\frac{\Delta\phi}{\mpl \mathcal{D}} \right)  \quad  
\eeq
and therefore validity of the EFT requires 
\beq
\frac{\left|\Delta m_\cT^2\right|}{m^2_\cT}\lesssim {\cal O}(1)\quad\Rightarrow\quad \frac{\Delta\phi}{\mpl}\lesssim \mathcal{D} \quad.
\eeq
Recent work supporting this conjecture was provided for example in refs.~\cite{Klaewer:2016kiy,Blumenhagen:2017cxt,Palti:2017elp,Grimm:2018ohb,Heidenreich:2018kpg}.

 \para{Swampland de Sitter Conjecture (SdSC) \cite{Ooguri:2016pdq,Freivogel:2016qwc,Brennan:2017rbf,Obied:2018sgi}} Motivated by recent discussions about finding de Sitter vacua in string theory compactifications (see e.g. \cite{Maldacena:2000mw,Bena:2014jaa,Kutasov:2015eba,Junghans:2016abx,Moritz:2017xto, Sethi:2017phn} and also \cite{Danielsson:2018ztv} for a recent review), it was recently conjectured that such vacua are incompatible with quantum gravity \cite{Obied:2018sgi}. In fact, the same conclusion was reached in ref.~\cite{Dvali:2013eja}   noting that de Sitter space  has a finite quantum breaking time. These observations lead to the conjecture that the inequality
 \beq
 \mpl | \nabla V | > c V \label{eq:conj2} \quad 
 \eeq
  holds at any point in moduli space, $\nabla V$ being the gradient of the potential. Again, $c$ is a positive order one constant that depends on details of the compactification. A weaker version of this condition was proposed in ref.~\cite{Obied:2018sgi}  following a similar line of reasoning to the SDC, stating that there exists an infinite tower of  states $\cT'$ whose masses scale as 
 \beq
 m^2_{\cT '} \sim \exp \left( -\frac{ cV }{\mpl | \nabla V |  } \right)  \quad  , 
 \eeq
 therefore becoming massless with the approach to a point in moduli space which parametrically violates eq.~\eqref{eq:conj2}.

\para{Consequences for inflation} When taken together, these conjectures have important consequences for inflation.   
The slow-roll parameter $\epsilon$ is related to the parameter $c$ in the SdSC by
\beq
\epsilon =\dfrac{M^2_P}{2}\left(\dfrac{\nabla V}{V}\right)^2 \quad \Rightarrow  \quad  c < \sqrt{2\epsilon } \ ,
\eeq
setting the requirement $ c < \sqrt{2} $ for single field slow-roll inflation.\footnote{We are demanding $\epsilon<1$ for slow-roll inflation to work, yet inflation just requires $-\dot H/H^2<1$. This is illustrated by single-field models, like DBI inflation~\cite{Alishahiha:2004eh}, where a higher order kinetic term allows for such an inflationary process to happen even if $\epsilon \nless 1$ while still being consistent with CMB bounds on non-Gaussianity.}

The SDC also constrains inflation since, if we assume $\epsilon$ to be monotonically increasing (which famously leads to the Lyth bound \cite{Lyth:1996im}), the field displacement can be related to the total number of e-folds $N_e$ as
\beq
N_e \sqrt{2\epsilon}  \lesssim \dfrac{\Delta \phi}{\mpl}  \, .
\eeq
We find therefore that the SDC implies  that $ N_e \sqrt{2\epsilon} \lesssim \mathcal{D}  $. Together with the previous constraint this gives
\beq
c N_e  < \sqrt{2\epsilon} N_e  \lesssim \mathcal{D} \ . \label{eq:constraint}
\eeq
Since in slow-roll single field inflation $r=16\epsilon$, these conditions can be translated to  bounds on the tensor-to-scalar ratio as
\beq
8c^2 < r \lesssim 8\frac{\mathcal{D}^2}{N_e^2}\quad. \label{eq:constr-2}
\eeq
In the rest of this letter we explore the limits of this bound on $r$ by providing arguments for the size of both $\mathcal{D}$ and $c$.


\para{Conjectures are parametric}  One of the possible consequences of eq.~\eqref{eq:constraint}, which was already observed in \cite{Agrawal:2018own,Achucarro:2018vey} (for other consequences see \cite{Garg:2018reu,Kehagias:2018uem}), is that the condition cannot be satisfied  if both $c$ and $\mathcal{D}$ are strictly of order one and $N_e \gtrsim 60$. This lead the authors of these papers to the conclusion that single field slow-roll inflation is incompatible with quantum gravity and the   proposal of multi-field  inflation as a possible way out  \cite{Achucarro:2018vey}.

In this letter we would like to relax this apparent tension between single field inflation and the above conjectures by noting that these conjectures should be read as \textit{parametric statements}.\footnote{For this purpose only we will take the conjectures at face value.} By this we mean that evidence supporting these conjectures 
 indicates that  $c$ and $\mathcal{D}$ don't need to be strictly of order unity. By relaxing these statements we will see it is indeed possible to satisfy eq.~\eqref{eq:constraint}, and we will convert this possibility into a rough lower bound on the tensor-to-scalar ratio $r$.

The conjectures above are mostly based on evidence from string theory compactifications  or EFT analysis. In the case of the  SDC conjecture, recent results supporting the conjecture come from string theory compactifications~\cite{Grimm:2018ohb} and EFT arguments \cite{Heidenreich:2018kpg}. These works show  that   parametrically large field-space displacements   arising when approaching a singularity in  moduli space give rise to   infinite towers of states becoming exponentially light. In fact, it was previously  observed in for example refs.~\cite{Kaloper:2011jz,McAllister:2014mpa,Baume:2016psm,Valenzuela:2016yny,Blumenhagen:2017cxt,Hebecker:2017lxm,Cicoli:2018tcq,Blumenhagen:2018nts}   that the validity of certain EFTs of large field inflation implies field range limits for the inflaton of order $\Delta\phi\sim \mpl\mathcal{D}$, where
\beq
\mathcal{D} \sim \left(\frac{m_h}{m_\ell}\right)^p> 1\quad,\quad p = 1,2, \dots\quad.
\eeq
Here $m_h$  denotes the mass of the lightest of the heavy fields, which were integrated out from the EFT,   and $m_\ell$ the mass of the light field relevant for the EFT. This behaviour is precisely what one should expect from  the general rules of decoupling in QFT and an important factor to be taken into account in any serious attempt to fix the constants in both conjectures.\footnote{Here we take the Wilsonian point of view, that any theory of quantum gravity including string theory should, as soon as its effective description as a QFT on curved background space-time emerges from the UV, reduce to a Wilsonian effective QFT at lower energy scales.} In this letter we do not seek to address  such an ambitious goal; we wish to emphasise that these constants are not  order unity and want to recall the relevance of hierarchies in relaxing the statements to parametric ones.

Moreover, we can argue purely in  terms of EFT arguments that establishing a finite mass hierarchy between the inflaton dynamics and the other sectors of quantum gravity allows for finitely large field excursions of the inflaton before noting the effect of the integrated out    fields. For this purpose, and following an example from~\cite{Dong:2010in}, we consider an effective model with two scalar fields -- a light one $\phi$ and a heavy one $\chi$ -- which captures many of the qualitative features of integrating out the heavy sectors present in UV complete models of inflation. We start with the potential
\beq
V(\phi, \chi ) =g^2\phi^2\chi^2+m_h^2\,(\chi-\chi_0)^2\quad ,
\eeq
where again $m_h$ denotes the heavy modulus  mass and  the light field $\phi$ has a mass 
$m_\ell=g\chi_0$ which depends on the vev of $\chi$.  We demand that $\chi_0\sim \mpl/2 $ \ to ensure that $\chi$ is heavy enough during inflation for all values of \ $\phi$  ($m_h/H> 3/2$).  The effective potential, upon integrating out the heavy field $\chi$, is:
\beq
V_{eff}(\phi ) =m_\ell^2\phi^2 \frac{1}{1+\frac{\phi^2}{M^2_{\rm P}\mathcal{D}^2}}\quad{\rm with }\quad \mathcal{D}=\dfrac{\chi_0}{\mpl}\frac{m_h}{m_\ell}\quad, 
\eeq
where $\mathcal{D}$ denotes the scale at which backreaction effects become important for large displacements of $\phi$. 
We   see   that enforcing a finite mass hierarchy $m_\ell<m_h $ provides a finitely large field range $\mpl\mathcal{D}$ over which backreaction remains small. Hence, the mass hierarchy directly implies  $\mathcal{D} > 1$, in agreement with e.g. ref.~\cite{Blumenhagen:2017cxt}. 

This EFT also allows us to connect with the SdSC. We can provide an estimate for the lower bound of $c$ by considering $|\nabla V|/V$ for the model above where it reduces to the equation 

\beq
   \mpl      \dfrac{|  V'_{eff}|}{V_{eff}}      =  \dfrac{2\mpl}{\phi} - \dfrac{   2 \phi}{\mpl \mathcal{D}^2}+\dots  \quad .
\eeq
Here the first term arises at tree level and the second one comes from integrating out the heavy field. Comparing the size of the tree level term to the magnitude of the heavy field contribution, we see that they never shrink below order $1/{\cal D}\sim m_\ell/m_h$ simultaneously. Hence, restricting ourselves to the regime of validity of the EFT given by the  SDC, this provides a lower bound on the ratio such that  $ c \sim m_\ell/m_h$. As it happened in the case of $\mathcal{D} $ before, we find that   imposing a hierarchy prevents $c$ from having to be order unity.  

\begin{figure}[t]
  \includegraphics[width=0.25\textwidth]{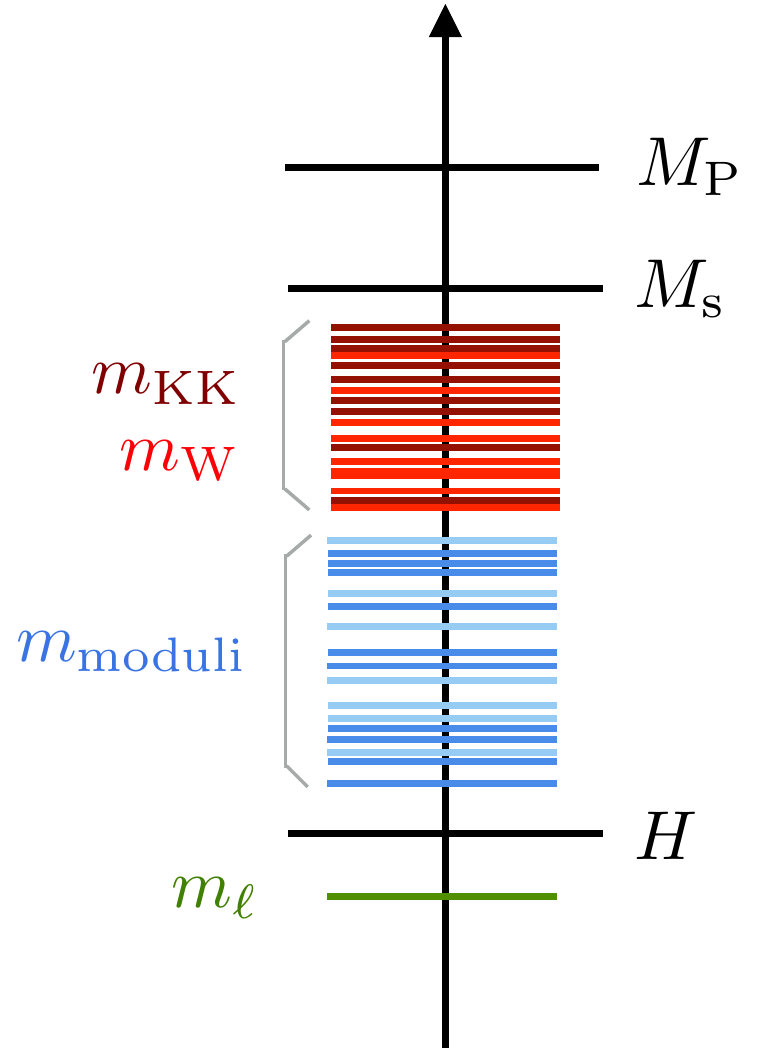}
  \caption{Mass spectrum in string theory. Generically we find little space to fit the masses of KK modes, winding modes and other moduli masses parametrically above $m_{\ell}$. The NLM with mass $m_h$ will belong to this last class. The light field with mass $m_l\lesssim H$ is the inflaton. }
   \label{fig:masses}
\end{figure}

Hence, we find that in the limit where the EFT starts to \textit{feel} the existence of the heavy field,  $c\sim 1/{\cal D}\sim m_\ell/m_h<1$. Based on the underlying reasoning from decoupling, we expect it to be true much more generally that a finite suppression of $ m_\ell/m_h$ can arise from a finite mass hierarchy when integrating out heavy states. 

Of course, here we were dealing with a particular model and in more general configurations one should expect a more general expression such as
\beq
c\sim \left(\dfrac{m_\ell}{m_h}\right)^{p'} \quad , \quad p'=1,2, ...
\eeq
Moreover, other factors should also be taken into account for a general expression for $c$ (and $\mathcal{D}$), but from here we see that there is no reason to believe that these constants are strictly order one.

Having understood the behaviour of $c$ and ${\cal D}$ as resulting from  hierarchies, we now turn to estimating reasonable values based on embedding into string theory compactification and the resulting spectrum of states.

\para{Consequences for observations}  As discussed, the possibility of having constants $c$ and $\mathcal{D}$ not strictly of order one arises from the existence of hierarchies in the mass spectrum of the theory of quantum gravity which UV completes the EFT describing inflation.  In the case of string theory, the heavy fields  could be string states, KK-modes, winding modes or moduli. The resulting hierarchies are
\beq
m_\ell \lesssim H < m_h \lesssim m_{\rm KK} \ , \ m_{\rm W}<M_{\rm s}<\mpl \ 
\eeq
where the relevant heavy mass for the ratio $m_{h}/m_{\ell}$ refers to the lightest of the heavy states. Fig.~\ref{fig:masses} illustrates the generic situation that the gap to the lightest of the heavy states with mass $m_h$, the `next-to-lightest modulus' (NLM) isn't very large. In other words, it is hard to find $\mathcal{D}$ parametrically large and $c$ parametrically small, in agreement with both SDC and SdSC. The behaviour of known examples (see e.g.~\cite{McAllister:2008hb,McAllister:2014mpa,Hebecker:2014kva,Bielleman:2016olv,Blumenhagen:2017cxt,Landete:2017amp}) of string compactification is consistent with 
\beq
\mathcal{D} \sim \dfrac{1}{c }\sim\dfrac{m_h}{m_\ell }\sim 10-100 \quad.
\eeq
\begin{figure}[t!]
  \includegraphics[width=0.45\textwidth]{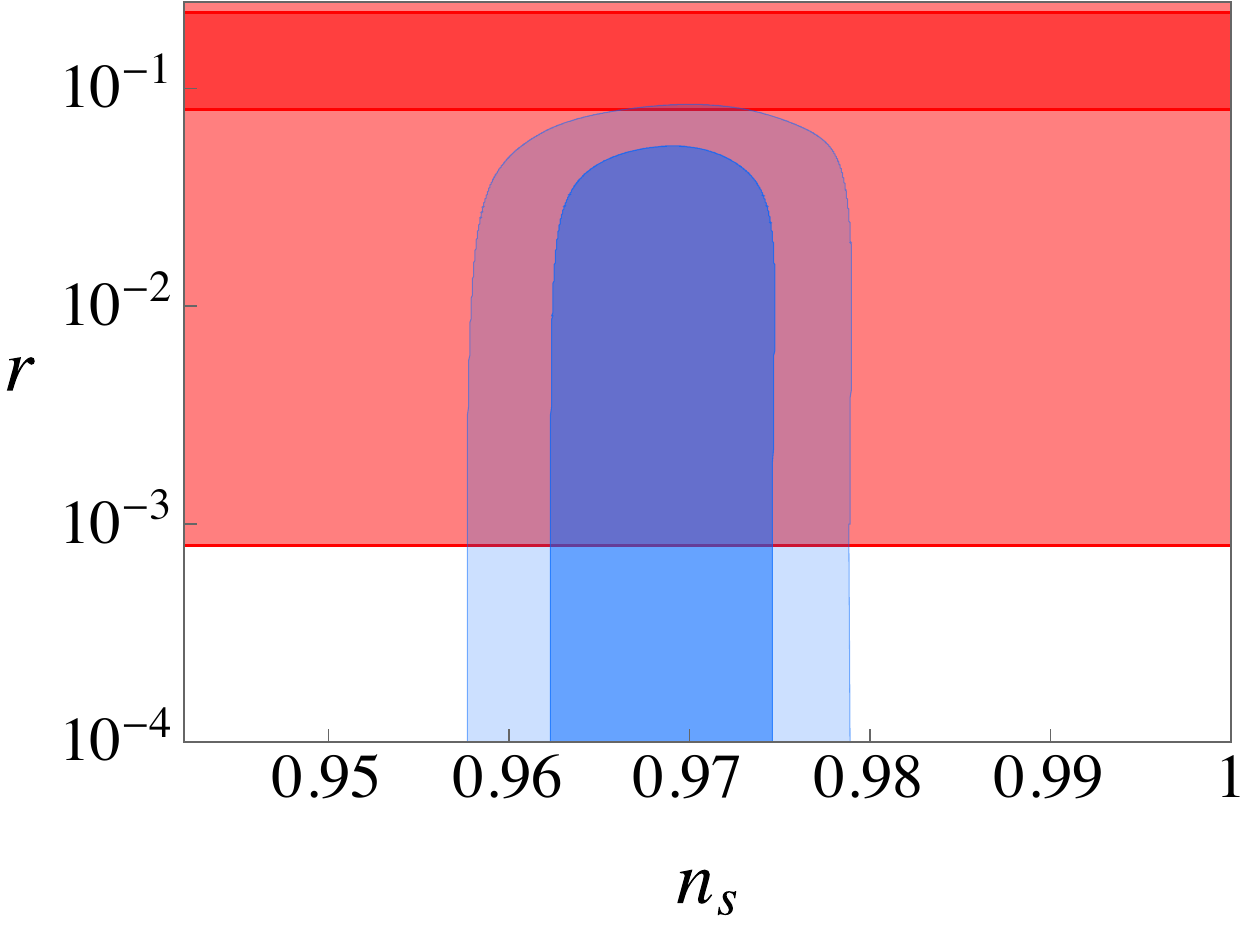}
  \caption{Comparison of swampland constraints (SDC + SdSC) and combined observational constraints from Planck and BICEP2/Keck \cite{Array:2015xqh}. Taking $N_e = 60$ and $\mathcal{D} = 1/c = m_{h}/m_{\ell}$, the darker red region shows the bounds from a NLM mass hierarchy of $m_{h}/m_{\ell} =10$, which gives $0.08 \leq r \leq 0.22$. We see that this hierarchy is already under some tension even with current constraints. For a mass hierarchy of $m_{h}/m_{\ell} =100$, the swampland constraints are much broader $0.0008 \leq r \leq 22$ (the upper bound is not shown). Nevertheless, even for this large hierarchy, it will be possible to probe the lower bound with future observational surveys.}
\label{fig:moneyplot}
\end{figure}

We use this as a reasonable estimate for the magnitude which the hierarchy between the NLM and the inflaton can take. This translates into bounds on the tensor-to-scalar ratio displayed in fig.~\ref{fig:moneyplot}:
\begin{align}
\frac{m_h}{m_{\ell}}= 10 \quad &\rightarrow \quad 0.1 \ \lesssim \ r \ \lesssim  \ 0.2\\ \nonumber
\frac{m_h}{m_{\ell}}= 100 \quad &\rightarrow \quad 10^{-3} \ \lesssim \ r \ \lesssim  \ 20 \ .
\end{align}
We see that in both cases the upper bounds are already less constraining than existing observational data \cite{Array:2015xqh}.  The lower bounds provide a \emph{ball park level expectation of the tensor-to-scalar ratio} from general compatibility conditions of embedding inflation into quantum gravity.

\para{Acknowledgements} We are grateful to Jakob Moritz, Eva Silverstein and Irene Valenzuela for useful comments. JF, AR and AW are supported by the ERC Consolidator Grant STRINGFLATION under the HORIZON 2020 grant agreement no.
647995. MD is supported by the German Science Foundation (DFG) within the Collaborative
Research Centre 676 Particles, Strings and the Early Universe.

\bibliography{references}

\end{document}